\documentclass[twocolumn,tighten,times]{aastex63} 

\usepackage{amsmath} 
\usepackage{amssymb}
\usepackage{afterpage}
\extrafloats{100}

\received{January 29, 2021}
\revised{March 17, 2021}
\accepted{March 22, 2021}

\newcommand{\galprop}{\textsc{GalProp}}
\newcommand{\helmod}{\textsc{HelMod}}
\newcommand{\fermilat}{\emph{Fermi}-LAT}

\shorttitle{Low-energy excess in cosmic ray iron}
\shortauthors{Boschini et al.}

\begin{document}

\title{
A discovery of a low-energy excess in cosmic-ray iron: an evidence of the past supernova activity in the Local Bubble
}

\author[0000-0002-6401-0457]{M.~J.~Boschini}
\affiliation{INFN, Milano-Bicocca, Milano, Italy}
\affiliation{CINECA, Segrate, Milano, Italy}

\author[0000-0002-7669-0859]{S.~{Della~Torre}}
\affiliation{INFN, Milano-Bicocca, Milano, Italy}

\author[0000-0003-3884-0905]{M.~Gervasi}
\affiliation{INFN, Milano-Bicocca, Milano, Italy}
\affiliation{Physics Department, University of Milano-Bicocca, Milano, Italy}

\author[0000-0003-1942-8587]{D.~Grandi}
\affiliation{INFN, Milano-Bicocca, Milano, Italy}
\affiliation{Physics Department, University of Milano-Bicocca, Milano, Italy}

\author[0000-0003-1458-7036]{G.~J\'{o}hannesson} 
\affiliation{Science Institute, University of Iceland, Dunhaga 3, IS-107 Reykjavik, Iceland}
\affiliation{NORDITA,  Roslagstullsbacken 23, 106 91 Stockholm, Sweden}

\author[0000-0002-2168-9447]{G.~{La~Vacca}}
\affiliation{INFN, Milano-Bicocca, Milano, Italy}
\affiliation{Physics Department, University of Milano-Bicocca, Milano, Italy}

\author[0000-0002-3729-7608]{N.~Masi}
\affiliation{INFN, Bologna, Italy}
\affiliation{Physics Department, University of Bologna, Bologna, Italy}

\author[0000-0001-6141-458X]{I.~V.~Moskalenko} 
\affiliation{Hansen Experimental Physics Laboratory, Stanford University, Stanford, CA 94305}
\affiliation{Kavli Institute for Particle Astrophysics and Cosmology, Stanford University, Stanford, CA 94305}

\author{S.~Pensotti}
\affiliation{INFN, Milano-Bicocca, Milano, Italy}
\affiliation{Physics Department, University of Milano-Bicocca, Milano, Italy}

\author[0000-0002-2621-4440]{T.~A.~Porter} 
\affiliation{Hansen Experimental Physics Laboratory, Stanford University, Stanford, CA 94305}
\affiliation{Kavli Institute for Particle Astrophysics and Cosmology, Stanford University, Stanford, CA 94305}

\author{L.~Quadrani}
\affiliation{INFN, Bologna, Italy}
\affiliation{Physics Department, University of Bologna, Bologna, Italy}

\author[0000-0002-1990-4283]{P.~G.~Rancoita}
\affiliation{INFN, Milano-Bicocca, Milano, Italy}

\author[0000-0002-7378-6353]{D.~Rozza}
\affiliation{INFN, Milano-Bicocca, Milano, Italy}
\affiliation{Physics Department, University of Milano-Bicocca, Milano, Italy}

\author[0000-0002-9344-6305]{M.~Tacconi}
\affiliation{INFN, Milano-Bicocca, Milano, Italy}
\affiliation{Physics Department, University of Milano-Bicocca, Milano, Italy}



\begin{abstract}

Since its launch, the Alpha Magnetic Spectrometer -- 02 (AMS-02) has delivered outstanding quality measurements of the spectra of cosmic-ray (CR) species, $\bar{p}$, $e^{\pm}$, and nuclei, $_1$H--\,$_8$O, $_{10}$Ne, $_{12}$Mg, $_{14}$Si, which resulted in a number of breakthroughs. One of the latest long awaited surprises is the spectrum of $_{26}$Fe just published by AMS-02. Because of the large fragmentation cross section and large ionization energy losses, most of CR iron at low energies is local, and may harbor some features associated with relatively recent supernova (SN) activity in the solar neighborhood. Our analysis of the new AMS-02 results together with Voyager 1 and ACE-CRIS data reveals an unexpected bump in the iron spectrum and in the Fe/He, Fe/O, and Fe/Si ratios at 1--2 GV, while a similar feature in the spectra of He, O, Si, and in their ratios is absent, hinting at a local source of low-energy CRs. The found excess extends the recent discoveries of radioactive $^{60}$Fe deposits in terrestrial and lunar samples, and in CRs.  We provide an updated local interstellar spectrum (LIS) of iron in the energy range from 1 MeV nucleon$^{-1}$ to $\sim$10 TeV nucleon$^{-1}$. Our calculations employ the \galprop{}--\helmod{} framework that is proved to be a reliable tool in deriving the LIS of CR $\bar{p}$, $e^{-}$, and nuclei $Z\le28$.

\end{abstract}



\section{Introduction} \label{Intro}

New era of precise astrophysical measurements has started about a decade ago with the launch of the Payload for Antimatter Matter Exploration and Light-nuclei Astrophysics \citep[PAMELA,][]{2007APh....27..296P,2014PhR...544..323A}. It was followed by a continuing series of launches of unique instrumentation, such as the {\it Fermi} Large Area Telescope \citep[\fermilat,][]{2009ApJ...697.1071A}, the Alpha Magnetic Spectrometer -- 02 \citep[AMS-02,][]{2013PhRvL.110n1102A}, NUCLEON experiment \citep{2019AdSpR..64.2546G, 2019AdSpR..64.2559G}, CALorimetric Electron Telescope -- \citep[CALET,][]{2019PhRvL.122r1102A, 2019AdSpR..64.2531T, 2020PAN....82..766M}, DArk Matter Particle Explorer mission -- \citep[DAMPE,][]{2017APh....95....6C,2017Natur.552...63D,eaax3793}, and Cosmic-Ray Energetics and Mass investigation -- \citep[ISS-CREAM,][]{2014AdSpR..53.1451S}. These experiments are operating in the high-energy and very-high-energy domains. 

Meanwhile, understanding the origin of cosmic rays (CRs) and our interstellar environment is impossible without connecting high energy measurements with data from low-energy experiments, such as the Cosmic Ray Isotope Spectrometer onboard of the Advanced Composition Explorer \citep[ACE-CRIS,][]{2018ApJ...865...69I, 2016Sci...352..677B} operating at the L1 Lagrange point for more than two decades, and Voyager~1,~2 spacecraft \citep{1977SSRv...21..355S}, the grandparents of the current instrumentation. The latter are providing unique data on the elemental spectra and composition at the {\it interstellar reaches} of the solar system \citep{2013Sci...341..150S, 2016ApJ...831...18C, 2019NatAs...3.1013S}, currently at 149 AU and 124 AU from the sun, correspondingly. 

These instruments and their experimental teams deliver CR data with unmatched precision and have not tired to surprise us with new-found features in the energy range that is deemed well-studied. One of the latest long awaited surprises is the spectrum of iron published by AMS-02 \citep{PhysRevLett.126.041104}. This is {\it an expected surprise} because CR iron is quite different from lighter species whose spectra have been recently published: $_1$H--\,$_8$O, $_{10}$Ne, $_{12}$Mg, $_{14}$Si \citep{2014PhRvL.113v1102A, 2015PhRvL.114q1103A, 2015PhRvL.115u1101A, 2016PhRvL.117i1103A, 2016PhRvL.117w1102A, 2017PhRvL.119y1101A, 2018PhRvL.120b1101A, 2018PhRvL.121e1103A, 2019PhRvL.122d1102A, 2019PhRvL.122j1101A, 2020PhRvL.124u1102A}. Because of the large fragmentation cross section of iron and large ionization energy losses at low energies, most of low-energy CR iron is local, coming from relatively recent supernova (SN) activity in the solar neighborhood, apart from lighter species that may come from more distant sources. The injection of iron is quite different from the injection of lighter species---it is injected into the interstellar space only during the SN explosion, while lighter elements are abundant in pre-supernova winds of massive stars. Besides, it has lower charge to mass ratio $(Z/A)_{\rm Fe} \approx 0.46$ compared to $(Z/A)_{\rm He, C, O, Si} \approx 1/2$ for lighter species and $(Z/A)_p=1$ for protons, and thus its acceleration in SN shock should be somewhat different from other nuclei. It is, therefore, natural to expect that the spectrum of CR iron should be somewhat different from other species, and we emphasized that numerous times \citep[e.g.,][]{2020ApJS..250...27B}.

In a recent paper \citep{2020ApJS..250...27B} we put forward a set of predicted spectra for all CR nuclei $_1$H--\,$_{28}$Ni, including those that are not yet published by AMS-02. Our predicted spectra for those species were based on Voyager~1 \citep{2016ApJ...831...18C} and ACE-CRIS data at low energies, while at high energies we used ATIC-2 \citep{2009BRASP..73..564P}, CREAM \citep{2008APh....30..133A}, NUCLEON \citep{2019AdSpR..64.2546G, 2019AdSpR..64.2559G}, CALET \citep{2019PhRvL.122r1102A}, and DAMPE \citep{eaax3793}.  In the intermediate range we used the HEAO-3-C2 data \citep{1990A&A...233...96E} that correspond to the aerogel counter of HEAO-3-C2 experiment. Our comparison with the published AMS-02 spectra of $_1$H--\,$_8$O, $_{10}$Ne, $_{12}$Mg, $_{14}$Si has shown that the HEAO-3-C2 data in the middle range from 2.65--10.6 GeV nucleon$^{-1}$, the so-called ``plateau'' corresponding to the aerogel counter, agree with the AMS-02 data quite well while at lower and higher energies the systematic deviations are large \citep[see a detailed discussion in][]{2020ApJS..250...27B}. However, the newly measured precise spectrum of iron just published by AMS-02 \citep{PhysRevLett.126.041104} is indeed harboring a surprise that provides an exciting opportunity to gain insight into the origin and evolution of the local interstellar medium.

In this paper we discuss the particularity of the measured spectrum of iron. Our calculations and interpretation employ the \galprop{}\footnote{Available from http://galprop.stanford.edu \label{galprop-site}}--\helmod{}\footnote{http://www.helmod.org/ \label{helmod-footnote}} framework that is proved to be a reliable tool in deriving the LIS of CR species \citep{2019HelMod, 2020ApJS..250...27B}.

\section{Calculations} \label{calcs}

\begin{deluxetable}{rlcc}[tb!]
	\def\arraystretch{0.9}
	\tablewidth{0mm}
	\tablecaption{Best-fit propagation parameters for {\it I}- and {\it P}-scenarios\label{tbl-prop}}
	\tablehead{
		\colhead{Parameter}& \multicolumn{1}{l}{Units}& \colhead{Best Value}& \colhead{Error} 
	}
	\startdata
	$z_h$ & kpc &4.0 &0.6\\
	$D_0 (R= 4\ {\rm GV})$ & cm$^{2}$ s$^{-1}$  & $4.3\times10^{28}$ &0.7\\
	$\delta$\tablenotemark{a} & &0.415 &0.025\\
	$V_{\rm Alf}$ & km s$^{-1}$ &30 &3\\
	$dV_{\rm conv}/dz$ & km s$^{-1}$ kpc$^{-1}$ & 9.8 &0.8
	\enddata
	\tablenotetext{a}{The {\it P}-scenario assumes a break in the diffusion coefficient with index $\delta_1=\delta$ below the break and index $\delta_2=0.15\pm 0.03$ above the break at $R=370\pm 25$ GV \citep[for details see][]{2020ApJ...889..167B}.}
\end{deluxetable}

\begin{deluxetable*}{@{}r@{}l rcl@{\hspace{20pt}}rcl@{\hspace{20pt}}rcl@{\hspace{20pt}}rcl@{\hspace{20pt}}r}[tb!]        \tabletypesize{\footnotesize}
        \def\arraystretch{1.1}
        \tablecolumns{15}
        \tablewidth{0mm}
        \tablecaption{The injection spectrum of iron \label{tbl-inject}}
        \tablehead{
                \multicolumn{2}{l}{}&
                \multicolumn{13}{c}{Spectral parameters}\\
                \cline{3-15}
                \multicolumn{2}{l}{Nucleus}&
                \multicolumn{1}{c}{\ $\gamma_0$} & \multicolumn{1}{c}{$R_0$ (GV)} & \multicolumn{1}{l}{$s_0$} &
                \multicolumn{1}{c}{$\gamma'$} & \multicolumn{1}{c}{$R'$ (GV)} & \multicolumn{1}{l}{$s'$} &
                \multicolumn{1}{c}{$\gamma_1$} & \multicolumn{1}{c}{$R_1$ (GV)} & \multicolumn{1}{l}{$s_1$} &
                \multicolumn{1}{c}{$\gamma_2$} & \multicolumn{1}{c}{$R_2$ (GV)} & \multicolumn{1}{l}{$s_2$} &
                \multicolumn{1}{c}{$\gamma_3$}  
                }
	\startdata
	Old $_{26}$ & Fe &
	0.27 & 1.04 & 0.18 & 
	\nodata & \nodata & \nodata & 
	1.99 & 7.00 & 0.20 & 
	2.51 & 355 & 0.17 &
	2.19\\
	New $_{26}$ & Fe &
	0.95 & 2.00 & 0.20 & 
	3.62 & 2.94 & 0.10 & 
	2.05 & 17.0 & 0.18 & 
	2.452 & 355 & 0.17 &
	2.23\\
	\enddata
  \tablecomments{The primary abundance of $^{56}$Fe is increased from 515 to 577 \citep[c.f.\ Tables 2, 3 in][]{2020ApJS..250...27B}, with relative abundances of isotopes of Fe after propagation tuned to ACE-CRIS data.}
\end{deluxetable*}

In this work we are using the same CR propagation model with distributed reacceleration and convection that was used in our previous analyses \citep[for more details see][]{2017ApJ...840..115B,  2018ApJ...854...94B, 2018ApJ...858...61B, 2020ApJS..250...27B, 2020ApJ...889..167B}. The latest versions of the \galprop{} code for Galactic propagation of CRs and the \helmod{} code for heliospheric propagation are described in detail in a recent paper by \citet{2020ApJS..250...27B}, see also references therein. 

The values of propagation parameters along with their confidence limits are derived from the best available CR data using the Markov Chain Monte Carlo (MCMC) routine. Five main propagation parameters, that affect the overall shape of CR spectra, were left free in the scan using \galprop{} running in the 2D mode: the Galactic halo half-width $z_h$, the normalization of the diffusion coefficient $D_0$ at the reference rigidity $R=4$ GV and the index of its rigidity dependence $\delta$, the Alfv\'en velocity $V_{\rm Alf}$, and the gradient of the convection velocity $dV_{\rm conv}/dz$ ($V_{\rm conv}=0$ in the plane, $z=0$). Their best-fit values tuned to the AMS-02 data are listed in Table~\ref{tbl-prop} and are the same as obtained in \citet{2020ApJS..250...27B}. The radial size of the Galaxy does not significantly affect the values of propagation parameters and was set to 20 kpc. Besides, we introduced a factor $\beta^\eta$ in the diffusion coefficient, where $\beta=v/c$, and $\eta$ was left free. The best fit value of $\eta=0.70$ improves the agreement at low energies, and slightly affects the choice of injection indices ${\gamma}_0$ an ${\gamma}_1$. A detailed discussion of the injection ({\it I}) and propagation ({\it P}) scenarios of the 350 GV break can be found in \citet{2012ApJ...752...68V} and \citet{2020ApJS..250...27B}.

The corresponding B/C ratio also remains the same \citep[see Fig.~4 in][]{2020ApJS..250...27B}, and compares well with all available measurements: Voyager~1 \citep{2016ApJ...831...18C}, ACE-CRIS\footnote{http://www.srl.caltech.edu/ACE/ASC/level2/cris\_l2desc.html}, AMS-02 \citep{2018PhRvL.120b1101A}, ATIC-2 \citep{2009BRASP..73..564P}, CREAM \citep{2008APh....30..133A, 2009ApJ...707..593A}, and NUCLEON \citep{2019AdSpR..64.2559G}. 

In this calculation we are tuning only the injection spectrum of iron to match the new data from AMS-02 at high energies and ACE-CRIS and Voyager~1 at low energies, the old and new injection parameters are shown in Table~\ref{tbl-inject}. Compared to the old injection spectrum we have to add another break at low rigidities at $R'=2.94$ GV and a very steep spectrum with index $\gamma'=3.62$ between $R_0=2.00$ GV and $R'=2.94$ GV. The obtained agreement with data is good, but the new Fe spectrum is quite different from what was expected based on the fit that employs the HEAO-3-C2 data \citep{2020ApJS..250...27B}.

\section{Results} \label{results}

\begin{figure}[tb!]
	\centering
	\includegraphics[width=0.47\textwidth]{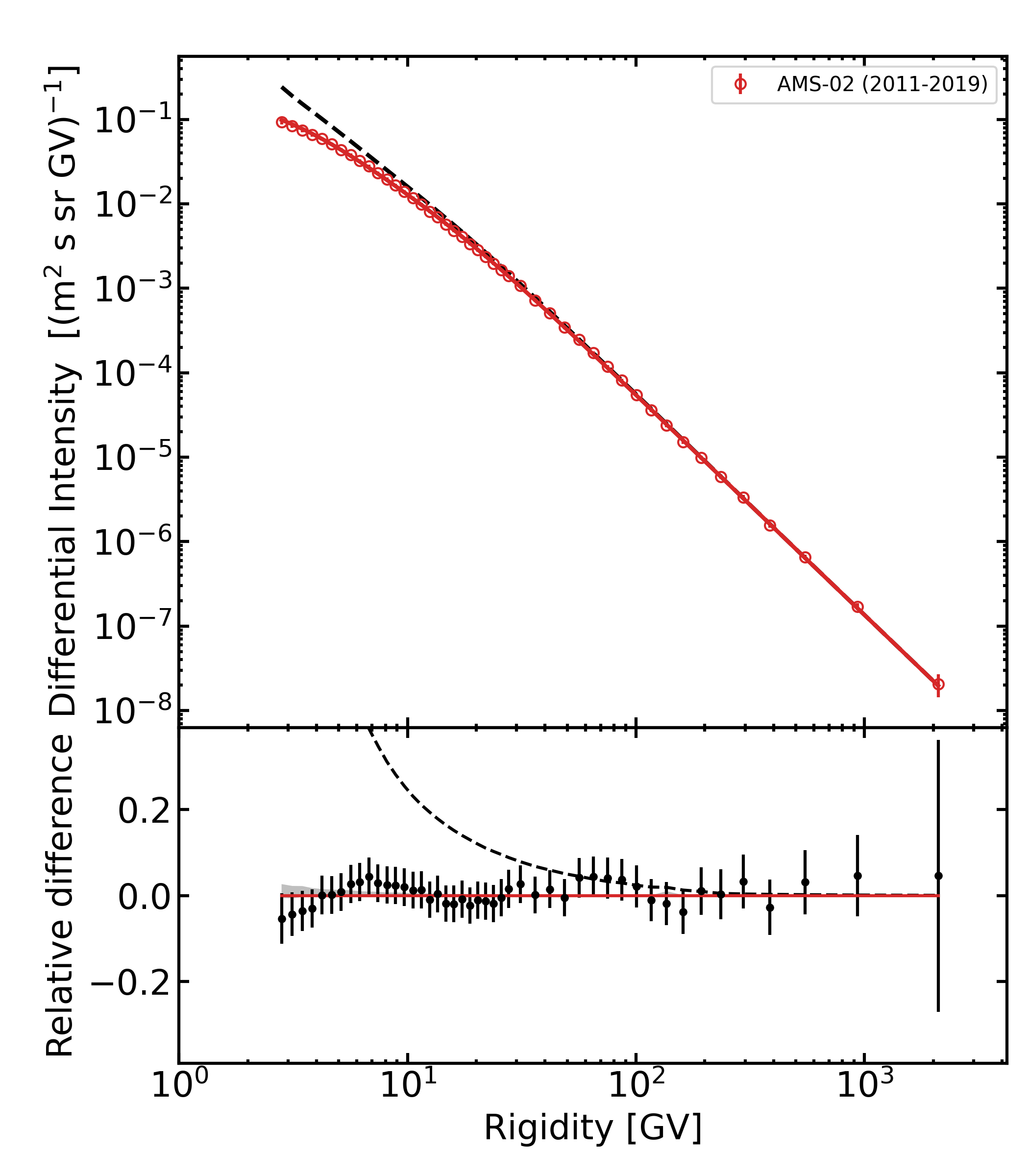}\vskip 5pt
	\includegraphics[width=0.47\textwidth]{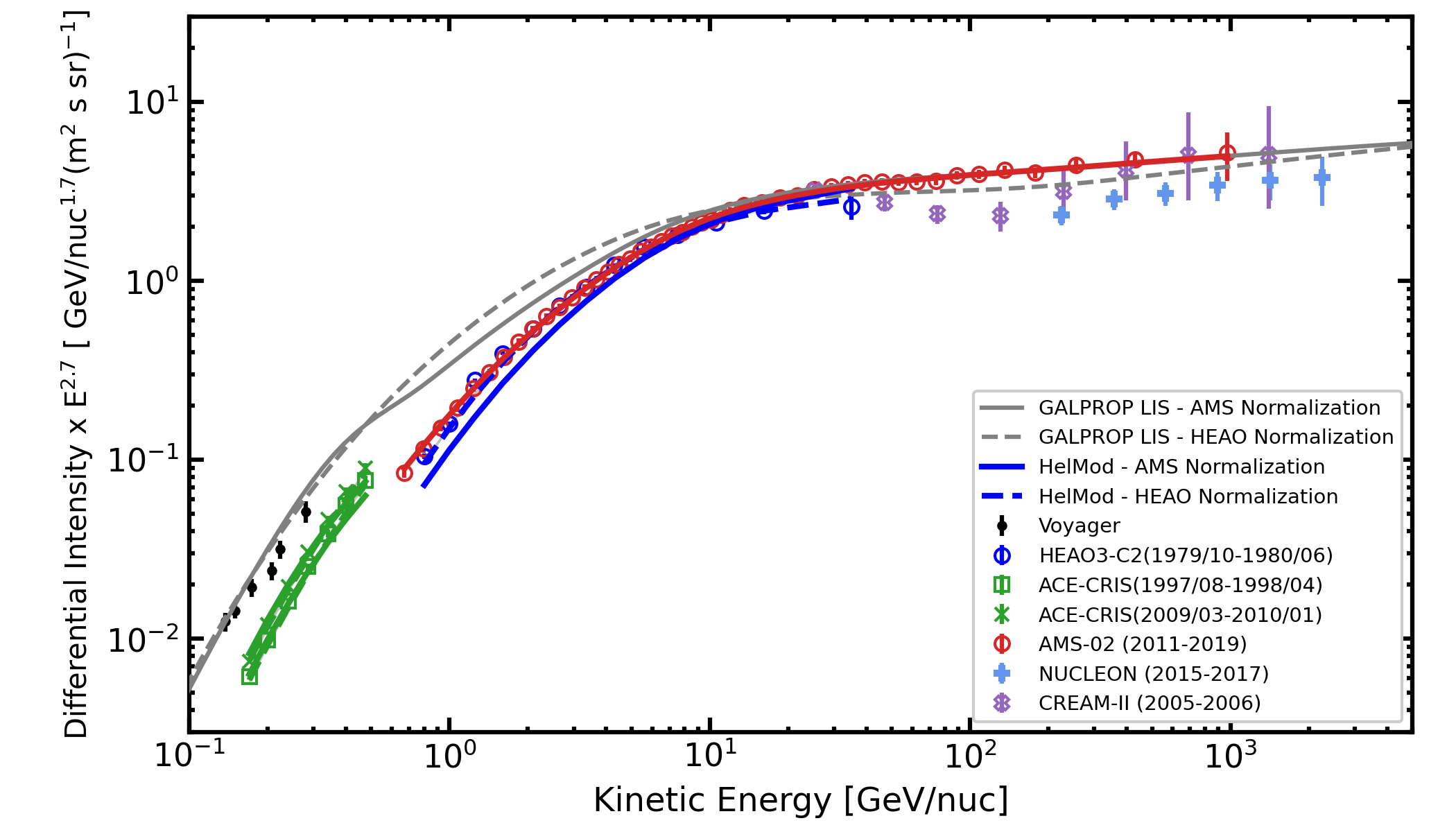}
	\caption{
A fit to the new spectrum of iron by AMS-02 \citep{PhysRevLett.126.041104}. In the top panel, only AMS-02 data are shown. The middle panel shows the quality of the fit, the relative difference between the calculations and the data set. 
The bottom panel shows two fits as compared to the data sets by HEAO-3-C2 \citep{1990A&A...233...96E} {\it or} by AMS-02 \citep{PhysRevLett.126.041104} in the intermediate range, while at low and very-high energies we use Voyager~1 \citep{2016ApJ...831...18C}, ACE-CRIS, CREAM \citep{2008APh....30..133A}, and NUCLEON \citep{2019AdSpR..64.2546G, 2019AdSpR..64.2559G} data. 
The gray lines show the LIS tuned to AMS-02 data (solid line) and the old LIS tuned to the ``plateau'' middle range of the HEAO-3-C2 data (dashed line). The red line shows the LIS modulated appropriately to the solar activity during the AMS-02 data taking, while solid blue line is the LIS derived from the AMS-02 data and modulated appropriately to the HEAO-3 flight. The two green lines correspond to two ACE-CRIS data taking periods, where the modulated spectra are almost independent on which data are used at high energies, AMS-02 or HEAO-3-C2,  dashed and solid lines are overlapping.
	}
	\label{fig:Fe-spec}
\end{figure}

\begin{figure}[tb!]
	\centering
	\includegraphics[width=0.45\textwidth]{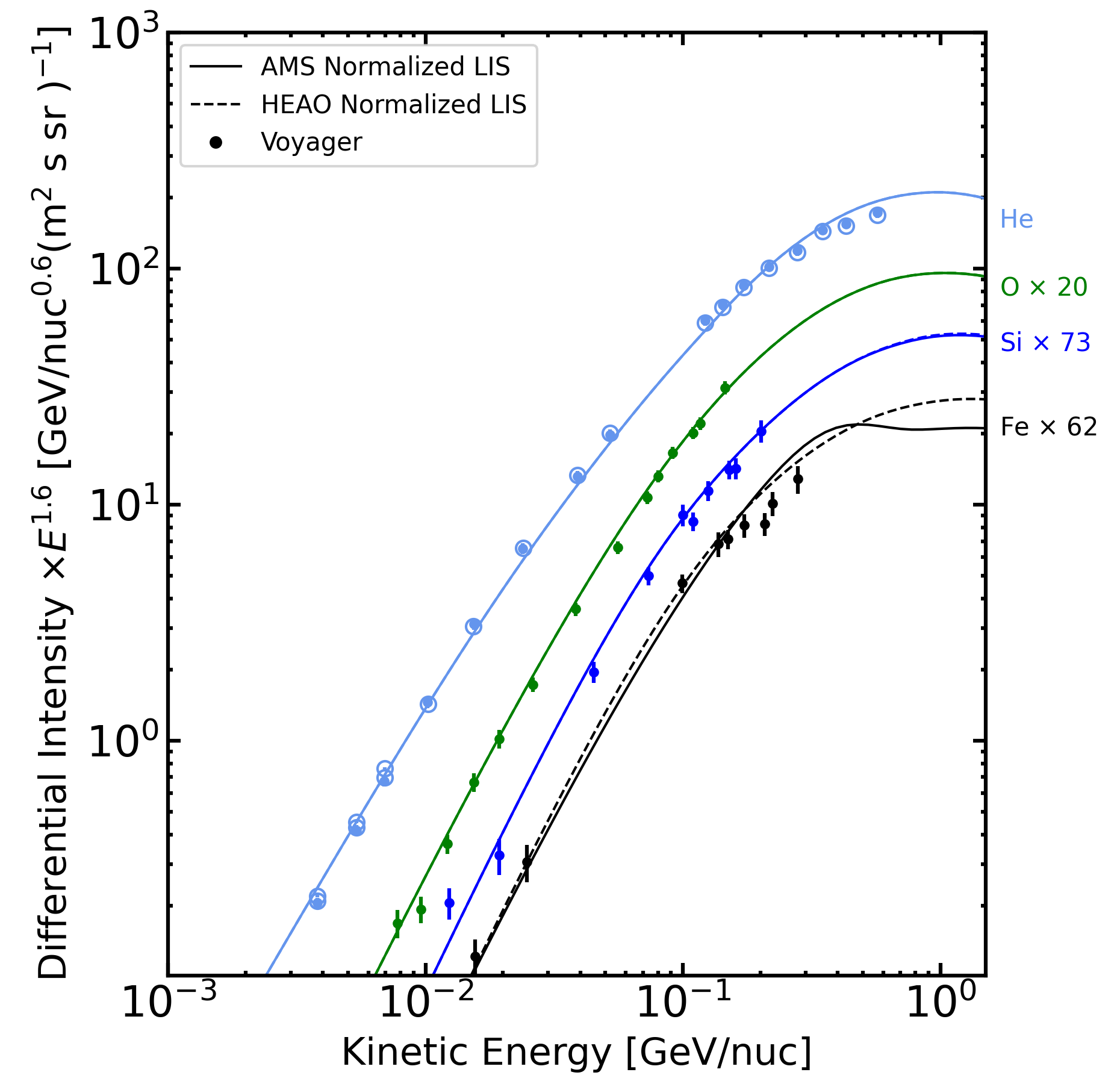}
	\caption{
The LIS of He, O, Si, and Fe compared to Voyager 1 data \citep{2016ApJ...831...18C}. The solid black line shows the updated Fe LIS tuned to AMS-02 data, while dashed line shows the previous LIS  \citep{2020ApJS..250...27B} based on the ``plateau'' middle range of the HEAO-3-C2 data.
	}
	\label{fig:Voyager}
\end{figure}

\begin{figure}[tb!]
	\includegraphics[width=0.425\textwidth]{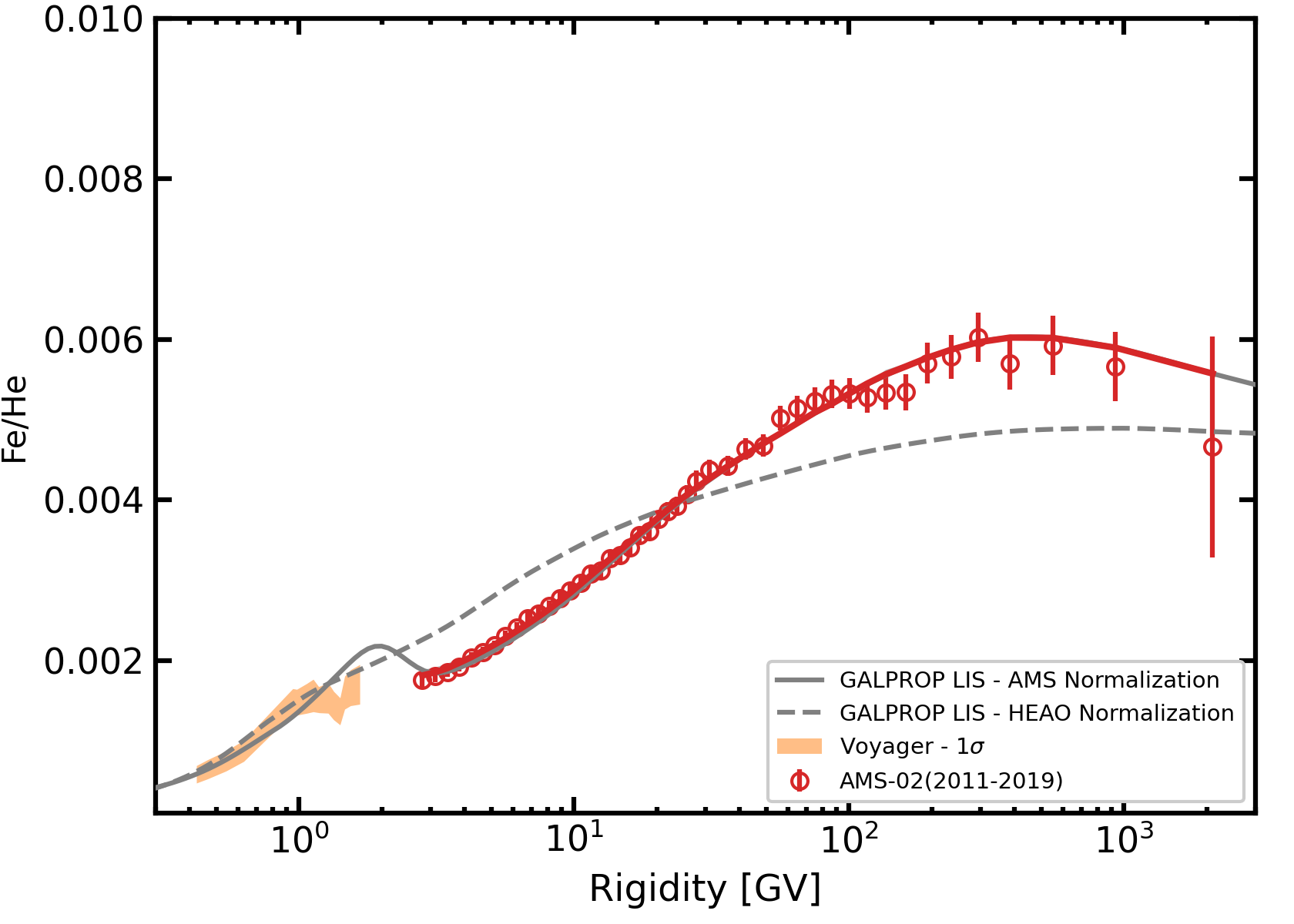}\\
	\includegraphics[width=0.425\textwidth]{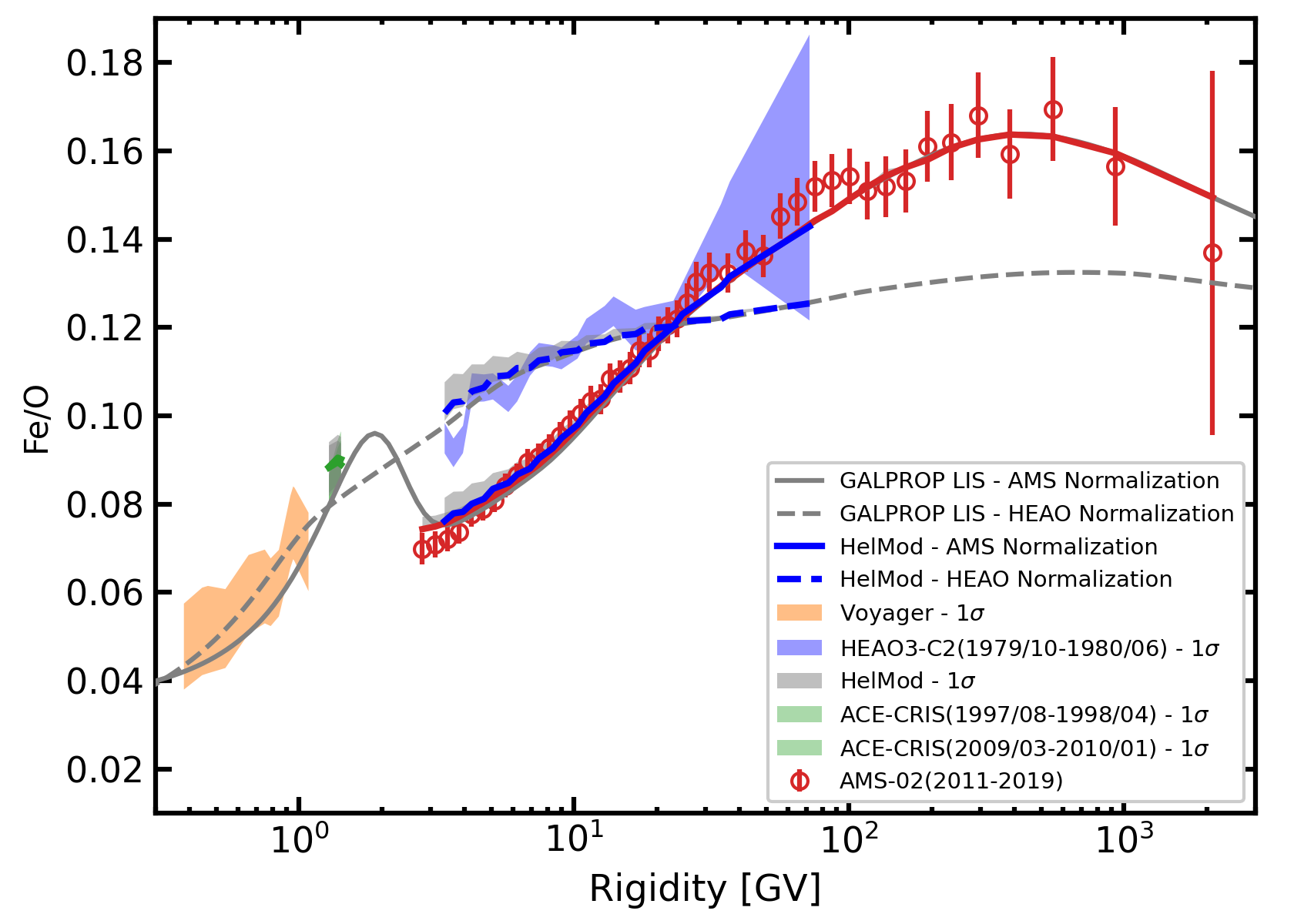}\\
	\includegraphics[width=0.425\textwidth]{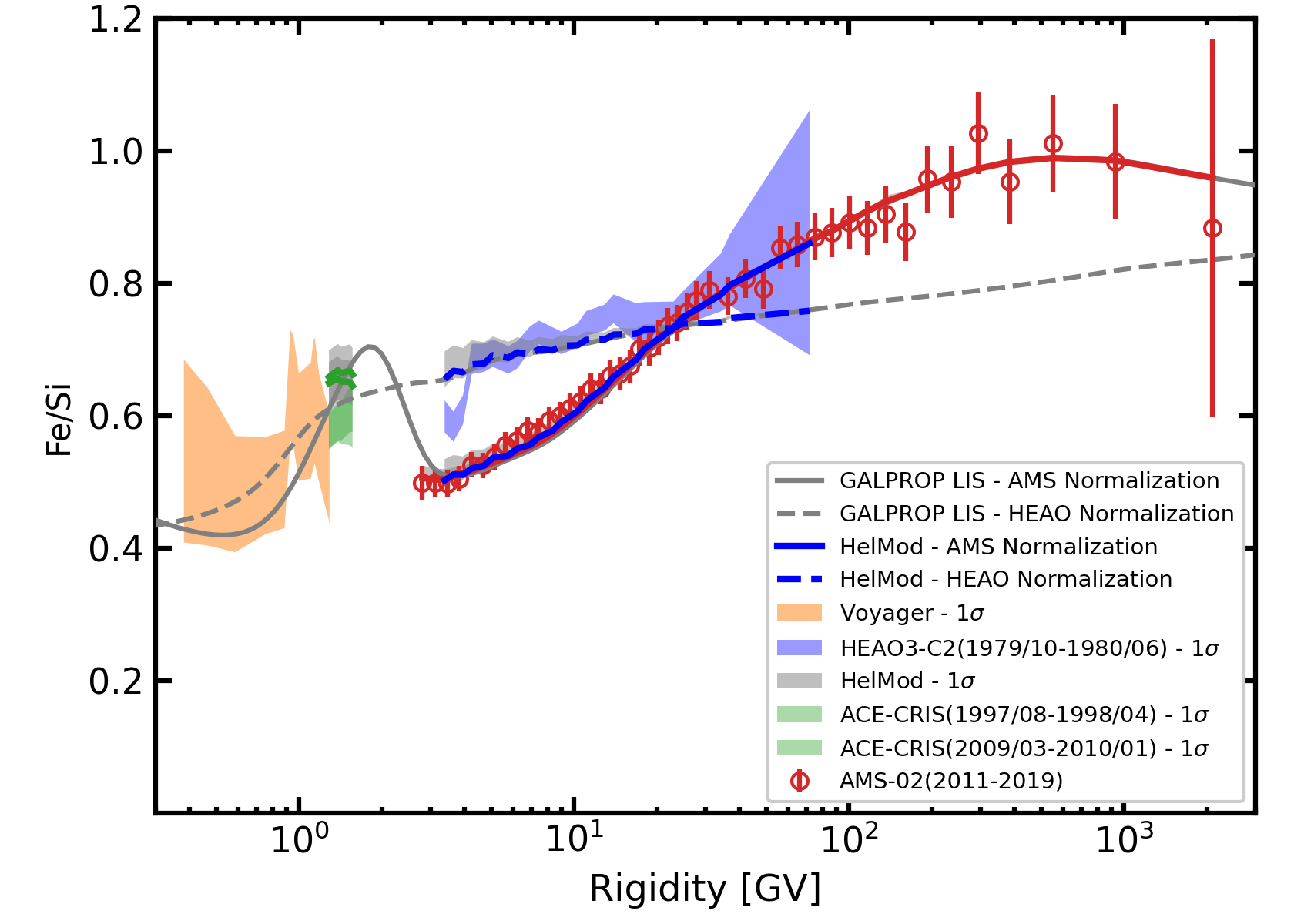}
	\caption{
The calculated ratios of primary species, Fe/He, Fe/O, Fe/Si, are compared with Voyager~1 \citep{2016ApJ...831...18C}, ACE-CRIS, and AMS-02 data \citep{2017PhRvL.119y1101A, PhysRevLett.126.041104, 2020PhRvL.124u1102A}. Also shown are HEAO-3-C2 data \citep{1990A&A...233...96E}.
The gray line shows the LIS ratios tuned to AMS-02 data, and the red line shows the modulated LIS ratios. The dashed gray line shows the LIS ratios tuned to HEAO-3-C2 data \citep{2020ApJS..250...27B} based on the ``plateau'' middle range of the HEAO-3-C2 data. For the Fe/He ratio, only Fe was tuned to the HEAO-3-C2 data. In all plots, the Voyager 1, ACE-CRIS, and HEAO-3-C2 data are converted from kinetic energy per nucleon to rigidity assuming $A/Z=2$. The shaded area shows the ratios modulated to the appropriate level (ACE-CRIS, HEAO-3-C2) with the width corresponding to 1$\sigma$ error.
	}
	\label{fig:Fe-ratios}
\end{figure}

\begin{figure}[tb!]
	\centering
	\includegraphics[width=0.425\textwidth]{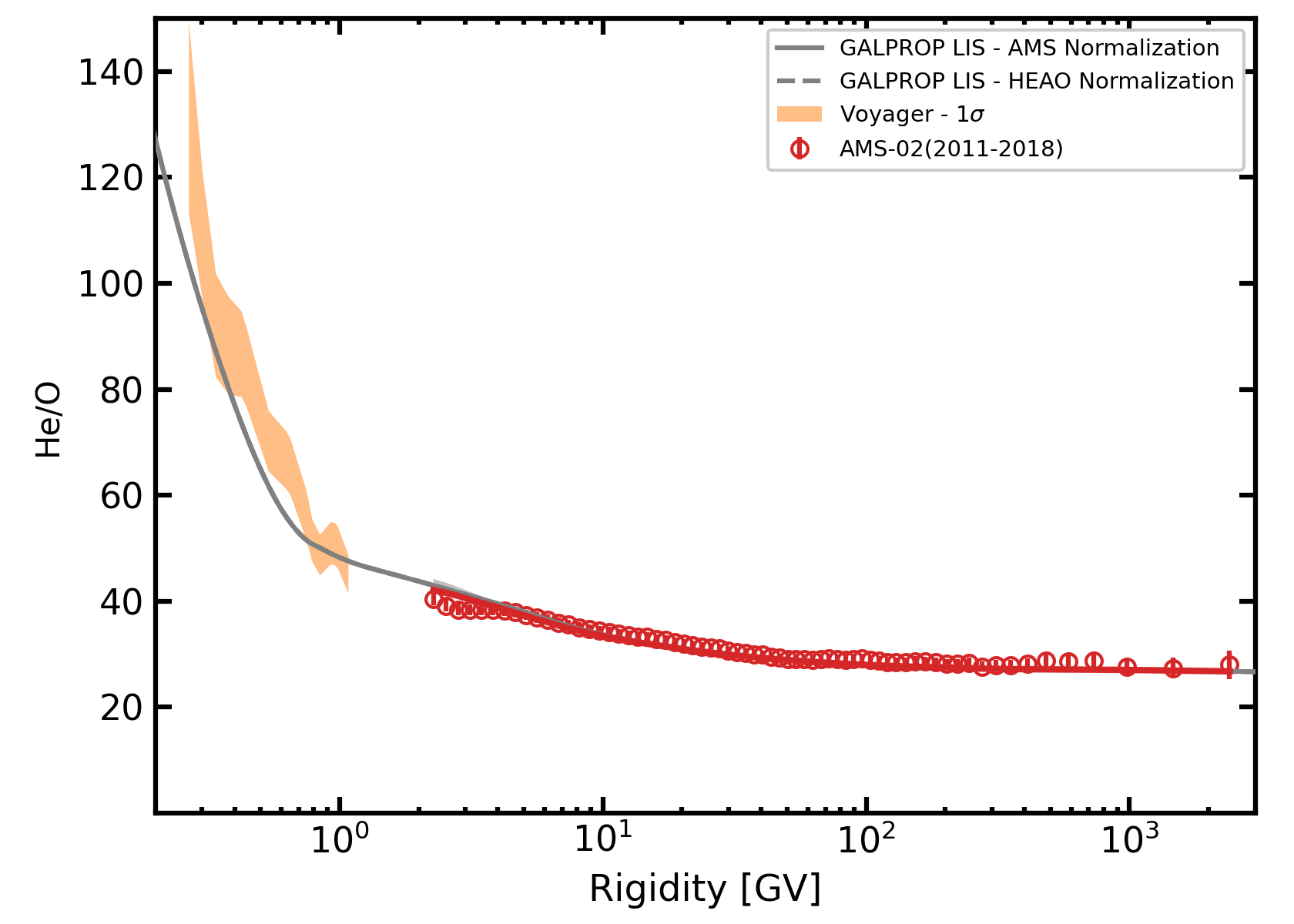}
	\includegraphics[width=0.425\textwidth]{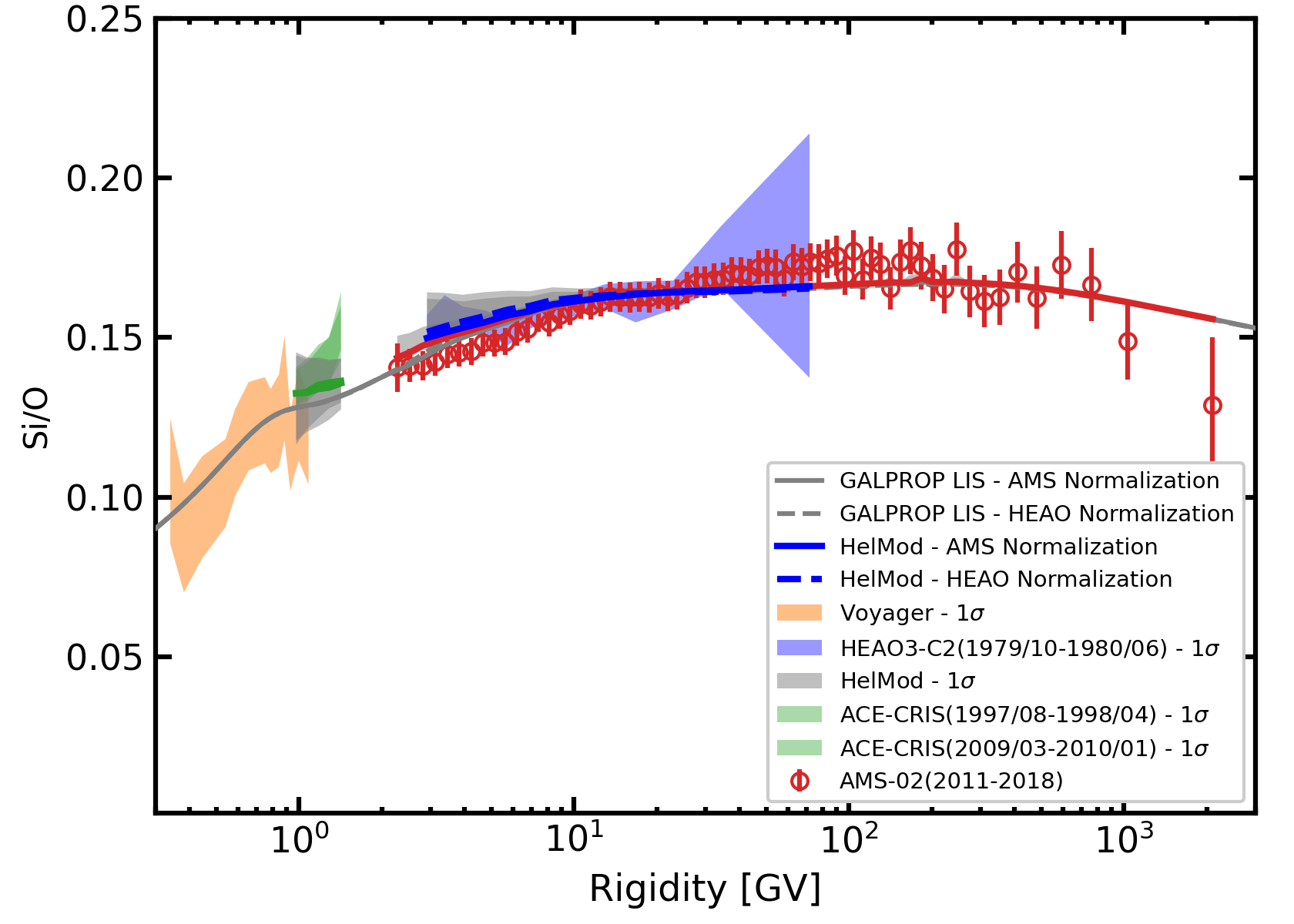}
	\caption{
The calculated ratios of primary species, He/O and Si/O, are compared with Voyager 1 \citep{2016ApJ...831...18C}, ACE-CRIS, HEAO-3-C2, and AMS-02 data \citep{2015PhRvL.115u1101A, 2017PhRvL.119y1101A, 2020PhRvL.124u1102A}. The Voyager 1, ACE-CRIS, and HEAO-3-C2 data are converted from kinetic energy per nucleon to rigidity assuming $A/Z=2$. 
	}
	\label{fig:O-ratios}
\end{figure}

Fig.~\ref{fig:Fe-spec} shows the fit to the iron spectrum by AMS-02 \citep{PhysRevLett.126.041104}. In the top panel, only AMS-02 data are shown. The quality of the fit is good with the residuals being $\la$5\% in the whole range (middle panel). The bottom panel shows two fits as compared to the data sets by HEAO-3-C2 \citep{1990A&A...233...96E} {\it or} by AMS-02 \citep{PhysRevLett.126.041104} in the intermediate range, while at low and very-high energies we use Voyager~1 \citep{2016ApJ...831...18C}, ACE-CRIS, CREAM \citep{2008APh....30..133A}, and NUCLEON \citep{2019AdSpR..64.2546G, 2019AdSpR..64.2559G} data. The gray lines show the LIS tuned to AMS-02 data (solid line) and the previous LIS tuned to the ``plateau'' middle range of the HEAO-3-C2 data (dashed line), while colored lines show the LIS modulated appropriately to the periods of ACE-CRIS, AMS-02, and HEAO-3 data taking. The solid blue line in Fig.~\ref{fig:Fe-spec} shows where the correct spectrum of HEAO-3-C2 should be if we assume the iron LIS based on AMS-02 data. One can see that matching the Voyager 1 and modulated ACE-CRIS data, on one hand, and modulated AMS-02 data, on the other hand, requires a sharp spectral steepening in the transition region of the LIS between $\sim$2 GV and $\sim$3 GV and a similar feature in the injection spectrum as well (see Table~\ref{tbl-inject}). 

Obvious is also the difference in the spectral shapes of He, O, Si, on one hand, and Fe, on the other hand (Fig.~\ref{fig:Voyager}). This difference becomes more obvious when one looks at the ratios. A comparison of the ratios of iron to lighter primary species, Fe/He, Fe/O, Fe/Si, with AMS-02 data shows excellent agreement in the rigidity range from 2 GV -- 2 TV (Fig.~\ref{fig:Fe-ratios}). If the ratios are extended to lower energies where they compare with Voyager 1 and ACE-CRIS data, one can see a clear bump at $\sim$1--2 GV. 
Therefore, the bumps appear in every ratio, Fe/He, Fe/O, Fe/Si, while a similar bump in the He/O and Si/O ratios is absent, see Fig.~\ref{fig:O-ratios}. 

We note that in the case of the iron flux, a normalization error in one of the instruments could result in the observed bump. However, a similar bump is also present in the Fe/He, Fe/O, Fe/Si ratios that makes a calibration error as the main cause for the bump rather unlikely.

Interestingly, the bump appears in the energy gap between ACE-CRIS and AMS-02, and is the result of tuning the Fe spectrum to all three data sets, Voyager~1, ACE-CRIS, AMS-02, each with an appropriate heliospheric modulation level. The anomaly would be unnoticed if the data of each instrument is taken separately. 
 
\begin{figure}[tbh!]
	\centering
	\includegraphics[width=0.425\textwidth]{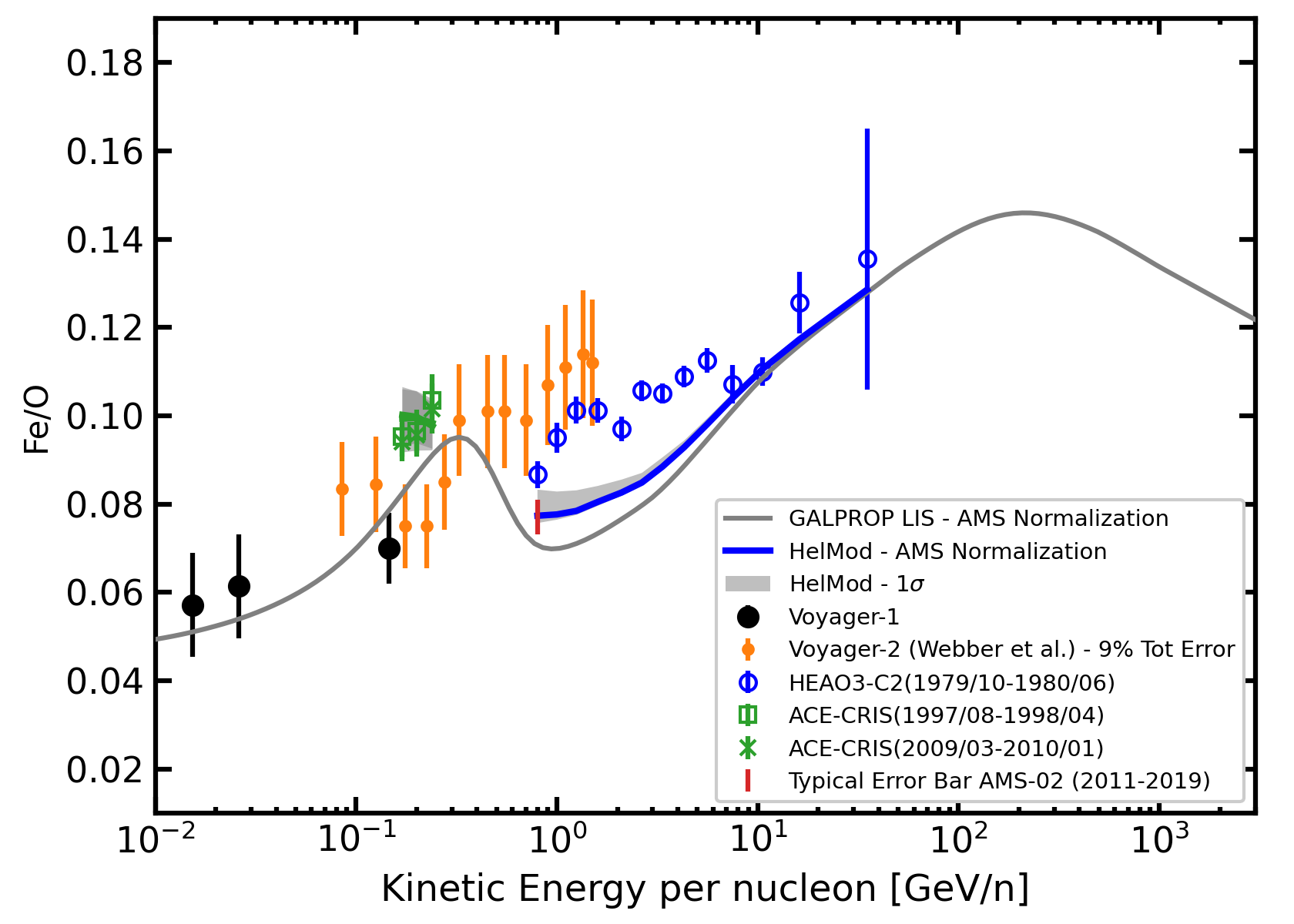}
	\caption{The Fe/O ratio vs.\ kinetic energy as measured by Voyager~1 \citep{2016ApJ...831...18C}, Voyager~2 \citep{2003ApJ...599..582W}, ACE-CRIS, and HEAO-3-C2 \citep{1990A&A...233...96E}. The plotted Voyager~2  error bars correspond to 9\% error. The gray line shows our derived LIS, and the blue line shows the modulated LIS at 1 au corresponding to the period of data taking by HEAO-3-C2. The shaded grey areas indicate 1$\sigma$ error for the \helmod{} calculations. The red bar shows typical error bars of AMS-02 Fe/O data points. 
	}
	\label{fig:FeO-ratio}
\end{figure}

So far we used the data taken either at 1 au (ACE-CRIS, AMS-02) or in the ISM (Voyager~1). Meanwhile, the analysis of Voyager~2 data that is using both penetrating and stopping particles in the High Energy Telescope (HET) extends the energy range from an upper energy limit of 100--300 MeV nucleon$^{-1}$ for stopping particles to over 1 GeV nucleon$^{-1}$ \citep{2003ApJ...599..582W}. This covers the energy gap between the ACE-CRIS and AMS-02 (see Fig.~\ref{fig:FeO-ratio}). 

The Voyager~2 data represent the cumulative spectra of the most abundant nuclei from Be to Fe that are derived for the so-called ``master time interval of $\sim$8 yr in duration covering both the 1986--1987--1988 and 1997--1998--1999 intensity maxima, as well as the years 1995 and 1996...'' \citep{2003ApJ...599..582W}. Both time intervals are covering the solar minima and the ascending parts of the solar cycles, while the instrument was moving with the speed of $\sim$2.9 au year$^{-1}$. 

The analysis of Voyager~2 data includes the standard energy range and the non-standard extended range. The standard energy ranges are defined for fully stopping particles. For oxygen and iron nuclei the standard ranges are 42--156 MeV nucleon$^{-1}$ and 82--300 MeV nucleon$^{-1}$, correspondingly, see Table 1 in the paper. The higher energy particles penetrate the entire HET telescope and require a non-standard analysis. Two different methods are used in the intervals 160--352 MeV nucleon$^{-1}$ and $>$450 MeV nucleon$^{-1}$ for oxygen, and 338--716 MeV nucleon$^{-1}$ and $>$700 MeV nucleon$^{-1}$ for iron. While in the lower energy interval the extension of the standard approach was used, the five highest energy data points from 700--1500 MeV nucleon$^{-1}$ were derived using a non-standard model-dependent approach that is not well documented. 

The extended energy range of Voyager 2 overlaps with HEAO-3-C2 instrument (Fig. \ref{fig:FeO-ratio}). The HEAO-3 instrument was flown at the near-Earth orbit at 1 au, and the published data were taken between October 17, 1979 and June 12, 1980. This period corresponds to the cycle 21 solar maximum conditions. One can see that the HEAO-3-C2 Fe/O ratio (vs.\ kinetic energy) taken during the solar {\it maximum at 1 au} is 1$\sigma$ {\it lower} than the Voyager 2 data taken during the solar {\it minima at 23 au and 54 au average radii.} Given that the Fe/O ratio {\it raises} with energy this demonstrates in the model-independent way that the Voyager 2 data above 700 MeV nucleon$^{-1}$ are too high. If the Voyager 2 Fe/O ratio above 700 MeV nucleon$^{-1}$ would be correct, it should go below the HEAO-3-C2 ratio. 

A large discrepancy in the Fe/O ratio between Voyager 2 and AMS-02 cannot be accounted for by the solar modulation and is likely a systematic error of the non-standard analysis of the Voyager 2 data. It is hard to specify the exact reasons, but we believe that the non-standard analysis technique and the correction for fragmentation of iron nuclei in the instrument are the main suspects. This assumption is supported by the fact that such analysis was never attempted again by the Voyager 1, 2 team. 

Despite the discrepancy above 700 MeV nucleon$^{-1}$, the analysis of lower energy data by Voyager~2 was performed in a more traditional way and agrees well with other experiments and with our derived LIS. In fact, it closes the energy gap between lower-energy instruments and AMS-02 thus offering a support to our conclusion on the low-energy excess in iron.

Meanwhile, the HEAO-3-C2 data are also too high based on a comparison with AMS-02. The discrepancies between HEAO-3-C2 and AMS-02 have been discussed in details in our previous paper \citep{2020ApJS..250...27B}. We extend the discussion in Section~\ref{heao3}.

\section{Discussion} \label{discussion}

The first ever accurate measurement of Fe spectrum in the rigidity range from 2 GV -- 2 TV \citep{PhysRevLett.126.041104} is a giant milestone. For the first time it allows us to analyze the spectra of primary species, He, O, Fe, whose atomic and mass numbers, fragmentation cross sections, and ionization energy losses are considerably different. The appearance of the bump in the Fe/He, Fe/O, and Fe/Si ratios, while lacking a similar feature in the He/O and Si/O ratios, implies a local source of low-energy CRs. Because of the fast ionization losses of iron nuclei in the ISM, CR iron from distant sources is suppressed and this gives the local source such a prominence. In the case of lighter CR nuclei, whose ionization losses are not that fast, the local component is mixed up with old Galactic CRs coming from distant sources and thus a prominent excess does not form. 

The likely source of the excess CR iron are the old SN remnants. In fact, the evidence of the past SN activity in the local ISM is abundant. The Local Bubble is a low density region of the size of $\sim$200 pc around the Sun filled with hot H\ {\sc i} gas that itself was formed in a series of SN explosions \citep[e.g.,][]{1999A&A...346..785S, 2011ARA&A..49..237F}. There are multiple reports of an excess of radioactive $^{60}$Fe found in the deep ocean sediments \citep{1999PhRvL..83...18K, 2004PhRvL..93q1103K, 2016PNAS..113.9232L, 2016Natur.532...69W},
in lunar regolith samples \citep{2009LPI....40.1129C, 2012LPI....43.1279F, 2014LPI....45.1778F}, and more recently in the Antarctic snow \citep{2019PhRvL.123g2701K}. Such deposits can be made by SN explosions in the solar neighborhood. Recent observation of $^{60}$Fe \citep[the half-life $\tau_{1/2}\sim2.6$ Myr,][]{2009PhRvL.103g2502R} in CRs by ACE-CRIS spacecraft \citep{2016Sci...352..677B} implies that the low-energy CRs from the most recent SN are still around.

It is hard to establish the number of SNe events and their exact timing, but it is clear that there could be several events during the last $\sim$10 Myr at distances of up to 100 parsecs \citep{2016Natur.532...69W}. The most recent SN events in the solar neighborhood were 1.5--3.2 Myr and 6.5--8.7 Myr ago \citep{2015ApJ...800...71F, 2016Natur.532...69W}. The measured spread (width) of the signal is $\sim$1.5 Myr \citep{2015ApJ...800...71F}, which is too long for a single event.  Each of these events could, in principle, consist of several consequent SN explosions separated by some 100 kyr, as an estimated time spread for a single SN, located at $\sim$100 pc from the Earth, is just $\sim$100--400 kyr and the travel time is $\sim$200 kyr. A detailed modeling by \citet{2016Natur.532...73B} indicates two SNe at distances 90--100 pc with the closest occurred 2.3 Myr ago and the second-closest exploded about 1.5 Myr ago, both with stellar masses of $\sim$9$M_\odot$. 
For discussions of other possible signatures of a local CR source see, e.g., \citet{PhysRevLett.121.251106} and \citet{Liu_2019}.

The new-found excess in the Fe spectrum below $\sim$2 GV is therefore falling in line with other excesses in $^{60}$Fe. However, this is the first time when the excess is found in the spectrum of the element that is dominated by {\it stable species}, most notably $^{56}$Fe. Such an excess has implications for the spectra of fragmentation products of iron. However, it is too early to make predictions for them as we have to await for a confirmation of similar features in the spectra of sub-Fe species.

Finally, we would like to comment on the validity of the propagation parameters derived from the lighter nuclei ratio, such as the B/C ratio, for the heavy nuclei in the iron group. A genuine probe of the propagation properties of the local ISM would be a comparison of the calculated (Sc+Ti+V)/Fe ratio aka sub-Fe/Fe ratio with its precise measurement. Currently, such measurements are only available from HEAO-3-C2 \citep{1990A&A...233...96E} in the energy range of interest and from ACE-CRIS and Voyager 1 \citep{2016ApJ...831...18C} at lower energies. Our detailed study shows that the calculated (Sc+Ti+V)/Fe ratio as well as individual spectra of each element, Sc, Ti, and V, are consistent with data \citep[see Figs.~8, 11, 12 in][]{2020ApJS..250...27B}. The relative source abundances of Sc and Ti, although non-negligible, are $\sim$3 times lower than their propagated abundances with Ti being the most abundant of all three \citep[see Fig.~7 in][]{2020ApJS..250...27B}.  Accordingly, less than $\sim$20\% of the total amount of Sc+Ti+V comes from the source and the rest is secondary, produced during the propagation. The relative source abundance of Ti is thus comparable with its relative abundance in the solar system derived from the Ti/Si and Ti/Fe ratios \citep{2003ApJ...591.1220L}. Therefore, the propagation parameters derived from the light nuclei ratio are not significantly different from the propagation properties of the local ISM. Relatively small deviations in the local propagation parameters, if exist, cannot be responsible for the observed feature. This supports our conclusion of its robustness. New accurate measurements of CR Sc, Ti, and V would be crucially important for the probes of the local ISM and the origin of the iron excess.

\section{HEAO-3-C2 data}\label{heao3}

In the previous paper we relied on the HEAO-3-C2 data \citep{1990A&A...233...96E} in the middle range, a so-called ``plateau'' region, to substitute the missing AMS-02 data. While it worked well for light and middle-range nuclei, it does not work so good for Fe -- as we have learned after AMS-02 data were published \citep{PhysRevLett.126.041104}. Surprisingly all HEAO-3-C2 data points for Fe overlap with AMS-02 data (Fig.~\ref{fig:Fe-spec}), but the solar modulation levels during the HEAO-3 flight and AMS-02 data taking are dramatically different. The former was launched on September 20, 1979 and ended on May 29, 1981, and took the data during the solar maximum conditions, while the AMS-02 data were taken from 2011--2019, i.e., through almost entire Cycle 24 where the solar activity was moderate even during the polarity reversal of the solar magnetic field. Therefore, the LIS derived from these two datasets are also quite different. The solid blue line in Fig.~\ref{fig:Fe-spec} shows where the correct spectrum of HEAO-3-C2 should be if we assume the iron LIS based on AMS-02 data. Here we discuss a possible reason for such a discrepancy. 

The AMS-02 experiment uses modern technology, its response was thoroughly simulated and tested, it also has several independent systems that allow for data cross checks. Therefore, it is rather unlikely to have a serious flaw. On the other hand, the HEAO-3-C2 experiment was built in 1970s, at the same time that Voyager 1, 2 were built. There was no prelaunch detector response simulations. Many systems went untested and the calibrations were done after the launch. In particular, the rigidity calibration was done using the geomagnetic field; the latter was lacking the detailed knowledge at that time \citep[for a discussion see][]{2020ApJS..250...27B}. Meanwhile, our analysis has shown that the middle range of the HEAO-3-C2 data, a so-called ``plateau,'' agrees well with AMS-02 data for light and medium nuclei. {\it There must be some reason why the situation with iron is different.}

While it is difficult to point to a single specific reason for such a difference, we think that the incorrect evaluation of the probability of nuclear fragmentation is the most likely. In the data analysis, the charge changing cross section was assumed to be energy-independent \citep{1990A&A...233...96E},
\begin{equation} \label{sigma}
\sigma=57.3 (A_t^{1/3}+A_p^{1/3}-0.83)^2,
\end{equation}
with parameters tuned to the measurements of charge changing reactions of $^{12}$C, $^{16}$O, and $^{56}$Fe beams with He target in the energy range from 540 MeV nucleon$^{-1}$ to 1600 MeV nucleon$^{-1}$ \citep{1988PhRvC..37.1490F}. Here $A_p$ and $A_t$ are the mass numbers of the projectile and target nuclei, correspondingly. Meanwhile, the average grammage for particles passing the instrument is 16 g cm$^2$ and the ``mean atomic number is $Z_t=30$.'' The mean atomic number of the material $Z_t$ is likely a typo and perhaps more correct would be to assume that it is the mean mass number $A_t=30$ rather than the atomic number. 

In any case, the instrumental grammage is significant and the loss of heavy CR nuclei due to the fragmentation was considerable. Therefore, the accuracy of a charge changing cross section is crucial for heavy species while it is less essential for light nuclei. The adopted scaling formula Eq.~(\ref{sigma}) was tuned to a rather light He target and {\it extrapolated} for $A_t=30$ target. We believe it is this correction for nuclear fragmentation for Fe that makes it different from lighter species and is the most likely reason for the disagreement with the AMS-02 data in the HEAO-3-C2 ``plateau'' energy range 2.65--10.6 GeV nucleon$^{-1}$. If our conclusion is correct, we should see a gradual increase in the discrepancy between the HEAO-3-C2 ``plateau'' data and AMS-02 data as the mass number increases from $^{14}$Si to $^{26}$Fe even though the agreement is good for all species from $^{4}$Be to $^{14}$Si.

We note that in the recent paper, \citet{2021arXiv210212576S} also discuss the low energy features in the combined iron spectrum from Voyager 1, ACE-CRIS, HEAO-3-C2, and AMS-02. Their paper does not offer any new insight into the origin of the observed features beyond an allusion that the likely reason for the discrepancy is an incorrect account for fragmentation of iron inside AMS-02. Such an interpretation is not viable because AMS-02 observes the Fe fragmentation inside the instrument and even measured the fragmentation cross section.

\section{Conclusion} \label{conclusion}

Using a combined data of AMS-02 \citep{PhysRevLett.126.041104}, ACE-CRIS, and Voyager 1 \citep{2016ApJ...831...18C}, we found an unexpected bump in the spectrum of CR iron and the Fe/He, Fe/O, and Fe/Si ratios, which is revealed as a sharp steepening in the range 2--3 GV necessary to connect ACE-CRIS and AMS-02 data. Meanwhile, a similar feature in the spectra of He, O, and Si and their ratios is absent. Taken independently, the data of each instrument do not show any feature. 

The new-found excess in the Fe spectrum below $\sim$2 GV is therefore falling in line with other excesses in $^{60}$Fe, which is likely connected to the past SN activity in the Local Bubble. However, this is the first time when the excess is found in the spectrum of the element that is dominated by {\it stable species}, most notably $^{56}$Fe. Such an excess has implications for the spectra of fragmentation products of iron. It is also important to measure the spectra of other heavy CR species, specifically in the middle rigidity domain $\la$10 GV, to see if a similar spectral feature is present. 

We are also discussing the reason for the discrepancy between AMS-02 and HEAO-3-C2 data. We argue that the likely reason is the incorrect estimate of fragmentation of CR iron nuclei inside the instrument. Given a significant instrumental grammage $\sim$16 g cm$^{-2}$ and its high mean mass number, the loss of heavy nuclei due to the fragmentation was considerable. Therefore, the accuracy of a charge changing cross section is crucial for heavy species while it is not so essential for light nuclei. If our conclusion is correct, we should see a gradual increase in the discrepancy between the HEAO-3-C2 ``plateau'' data and AMS-02 data as the mass number increases from $^{14}$Si to $^{26}$Fe even though the agreement is good for all species from $^{4}$Be to $^{14}$Si.


\acknowledgements
Special thanks to Pavol Bobik, Giuliano Boella, Karel Kudela, Marian Putis, and Mario Zannoni for their continuous support of the \helmod{} project and many useful suggestions. This work is supported by ASI (Agenzia Spaziale Italiana) through a contract ASI-INFN No.\ 2019-19-HH.0 and by ESA (European Space Agency) through a contract 4000116146/16/NL/HK. Igor V.\ Moskalenko and Troy A.\ Porter acknowledge support from NASA Grant No.~NNX17AB48G. We thank the ACE CRIS instrument team and the ACE Science Center for providing the ACE data. This research has made use of the SSDC Cosmic rays database
\citep{2017ICRC...35.1073D} and LPSC Database of Charged Cosmic Rays
\citep{2014A&A...569A..32M}. 

\bibliography{bibliography}

\appendix

\section{Analytical parameterization and numerical tables of the iron LIS}

Here we provide an analytical parameterization of the iron LIS: 
\begin{align}
    \label{eq:Fe}
F_{\rm Fe} (R)  = R^{-2.7}\times
&\begin{cases}
	\displaystyle a - b R + c R^2 - d R^3 + f \tanh\left[\tanh\left\{R G(-gR)\right\}\right]  - h G\left(i R^2\right),    &R\le 2.8\, {\rm GV}, \smallskip\\
	\displaystyle l - m R - n \tilde{R} + o \frac{\tilde{R}}{R^2} -\frac{p}{\tilde{R}}  + \frac{q}{R^2\tilde{R}} + r\log\left(s + R\right)   + \frac{t}{\tilde{R}\log\left(s + R\right)},  &R> 2.8\, {\rm GV},
\end{cases} 
\end{align}
where $R$ is the particle rigidity in GV, the values of the fitting parameters from $a$ to $t$ are given in Table \ref{Tbl-1-10}, and the $\tilde{R}$, and $G(x)$ functions are defined as: 
\begin{align}
	\tilde{R}&=\log{R},\nonumber\\
	G(x)&=e^{-x^2}.\nonumber
\end{align}
The analytical representation, Eq.~(\ref{eq:Fe}), is also complemented by numerical tables calculated for the {\it I}-scenario, which tabulate the LIS in rigidity $R$ (Table \ref{Tbl-IronLIS-Rigi}) and in kinetic energy $E_{\rm kin}$ per nucleon (Table \ref{Tbl-IronLIS-EKin}).

\begin{deluxetable*}{rlrlrlrl}[!hp]
	\tablecolumns{8}
	\tablewidth{0mm}
	\tablecaption{Parameters of the analytical fit to the iron LIS \label{Tbl-1-10}}
	\tablehead{
		\colhead{Param}     &  \colhead{Value}    & \colhead{Param}    & \colhead{Value}   &   \colhead{Param}     &  \colhead{Value}   & \colhead{Param}     &  \colhead{Value}
	}
	\startdata
	$a$\phantom{a}   & 2.8224e+0 & $g$\phantom{a}  & 2.3448e+0 & $m$\phantom{a}  & 5.9700e-8 & $q$\phantom{a} & 1.6412e+2 \\
	$b$\phantom{a}   & 3.3660e-1 & $h$\phantom{a}  & 2.8171e+0 & $n$\phantom{a}  & 5.6550e+0 & $r$\phantom{a} & 6.5309e+0 \\
	$c$\phantom{a}   & 6.1955e-1 & $i$\phantom{a}  & 3.8226e-1 & $o$\phantom{a}  & 2.4807e+2 & $s$\phantom{a} & 2.0316e+2 \\
	$d$\phantom{a}   & 1.2735e-1 & $l$\phantom{a}  & 2.6334e+1 & $p$\phantom{a}  & 1.3610e+2 & $t$\phantom{a} & 1.5300e+2 \\
	$f$\phantom{a}   & 2.3980e-1 & \nodata & \nodata & \nodata & \nodata & \nodata & \nodata\\
	\enddata
\end{deluxetable*}

\begin{deluxetable*}{cccccccccc}
	\tabletypesize{\footnotesize}
	\tablecolumns{10}
	\tablewidth{0mm}
	\tablecaption{$Z=26$ -- Iron LIS\label{Tbl-IronLIS-Rigi}}
	\tablehead{
		\colhead{Rigidity} & \colhead{Differential} &
		\colhead{Rigidity} & \colhead{Differential} &
		\colhead{Rigidity} & \colhead{Differential} &
		\colhead{Rigidity} & \colhead{Differential} &
		\colhead{Rigidity} & \colhead{Differential} \\ [-2ex] 
		\colhead{GV} & \colhead{intensity} &
		\colhead{GV} & \colhead{intensity} &
		\colhead{GV} & \colhead{intensity} &
		\colhead{GV} & \colhead{intensity} &
		\colhead{GV} & \colhead{intensity}  
	}
	\startdata
	9.299e-02 & 5.252e-03 & 7.508e-01 & 3.968e-01 & 1.039e+01 & 1.475e-02 & 5.430e+02 & 6.616e-07 & 3.413e+04 & 1.376e-11\\
	9.758e-02 & 5.891e-03 & 7.892e-01 & 4.239e-01 & 1.127e+01 & 1.234e-02 & 5.977e+02 & 5.152e-07 & 3.759e+04 & 1.069e-11\\
	1.024e-01 & 6.587e-03 & 8.297e-01 & 4.510e-01 & 1.224e+01 & 1.029e-02 & 6.580e+02 & 4.013e-07 & 4.139e+04 & 8.297e-12\\
	1.075e-01 & 7.365e-03 & 8.724e-01 & 4.780e-01 & 1.330e+01 & 8.542e-03 & 7.243e+02 & 3.126e-07 & 4.558e+04 & 6.441e-12\\
	1.128e-01 & 8.233e-03 & 9.175e-01 & 5.043e-01 & 1.448e+01 & 7.060e-03 & 7.974e+02 & 2.436e-07 & 5.019e+04 & 4.999e-12\\
	1.183e-01 & 9.204e-03 & 9.651e-01 & 5.300e-01 & 1.576e+01 & 5.812e-03 & 8.779e+02 & 1.898e-07 & 5.527e+04 & 3.880e-12\\
	1.242e-01 & 1.029e-02 & 1.015e+00 & 5.547e-01 & 1.718e+01 & 4.766e-03 & 9.665e+02 & 1.479e-07 & 6.086e+04 & 3.012e-12\\
	1.303e-01 & 1.150e-02 & 1.069e+00 & 5.778e-01 & 1.873e+01 & 3.889e-03 & 1.064e+03 & 1.152e-07 & 6.702e+04 & 2.337e-12\\
	1.368e-01 & 1.285e-02 & 1.125e+00 & 5.987e-01 & 2.045e+01 & 3.162e-03 & 1.172e+03 & 8.974e-08 & 7.380e+04 & 1.813e-12\\
	1.435e-01 & 1.435e-02 & 1.184e+00 & 6.164e-01 & 2.233e+01 & 2.561e-03 & 1.290e+03 & 6.989e-08 & 8.126e+04 & 1.407e-12\\
	1.506e-01 & 1.603e-02 & 1.248e+00 & 6.309e-01 & 2.440e+01 & 2.065e-03 & 1.420e+03 & 5.443e-08 & 8.948e+04 & 1.092e-12\\
	1.581e-01 & 1.790e-02 & 1.315e+00 & 6.415e-01 & 2.668e+01 & 1.657e-03 & 1.564e+03 & 4.238e-08 & 9.854e+04 & 8.468e-13\\
	1.659e-01 & 1.998e-02 & 1.386e+00 & 6.476e-01 & 2.920e+01 & 1.325e-03 & 1.722e+03 & 3.300e-08 & 1.085e+05 & 6.569e-13\\
	1.741e-01 & 2.230e-02 & 1.461e+00 & 6.486e-01 & 3.196e+01 & 1.057e-03 & 1.896e+03 & 2.570e-08 & 1.195e+05 & 5.095e-13\\
	1.827e-01 & 2.488e-02 & 1.542e+00 & 6.436e-01 & 3.500e+01 & 8.402e-04 & 2.087e+03 & 2.001e-08 & 1.316e+05 & 3.952e-13\\
	1.918e-01 & 2.775e-02 & 1.627e+00 & 6.317e-01 & 3.835e+01 & 6.665e-04 & 2.298e+03 & 1.559e-08 & 1.449e+05 & 3.065e-13\\
	2.013e-01 & 3.093e-02 & 1.718e+00 & 6.120e-01 & 4.204e+01 & 5.275e-04 & 2.531e+03 & 1.215e-08 & 1.596e+05 & 2.377e-13\\
	2.112e-01 & 3.447e-02 & 1.816e+00 & 5.839e-01 & 4.610e+01 & 4.166e-04 & 2.787e+03 & 9.464e-09 & 1.757e+05 & 1.843e-13\\
	2.217e-01 & 3.839e-02 & 1.919e+00 & 5.472e-01 & 5.057e+01 & 3.283e-04 & 3.068e+03 & 7.374e-09 & 1.935e+05 & 1.429e-13\\
	2.327e-01 & 4.273e-02 & 2.031e+00 & 5.031e-01 & 5.549e+01 & 2.583e-04 & 3.379e+03 & 5.744e-09 & 2.131e+05 & 1.108e-13\\
	2.442e-01 & 4.754e-02 & 2.150e+00 & 4.539e-01 & 6.091e+01 & 2.029e-04 & 3.720e+03 & 4.474e-09 & \nodata &\nodata\\
	2.563e-01 & 5.286e-02 & 2.277e+00 & 4.025e-01 & 6.688e+01 & 1.592e-04 & 4.097e+03 & 3.485e-09 & \nodata &\nodata\\
	2.690e-01 & 5.873e-02 & 2.414e+00 & 3.521e-01 & 7.344e+01 & 1.247e-04 & 4.511e+03 & 2.713e-09 & \nodata &\nodata\\
	2.823e-01 & 6.521e-02 & 2.561e+00 & 3.051e-01 & 8.068e+01 & 9.766e-05 & 4.967e+03 & 2.113e-09 & \nodata &\nodata\\
	2.964e-01 & 7.234e-02 & 2.720e+00 & 2.633e-01 & 8.864e+01 & 7.640e-05 & 5.470e+03 & 1.645e-09 & \nodata &\nodata\\
	3.111e-01 & 8.019e-02 & 2.891e+00 & 2.274e-01 & 9.741e+01 & 5.972e-05 & 6.023e+03 & 1.280e-09 & \nodata &\nodata\\
	3.265e-01 & 8.881e-02 & 3.075e+00 & 1.974e-01 & 1.071e+02 & 4.665e-05 & 6.632e+03 & 9.964e-10 & \nodata &\nodata\\
	3.428e-01 & 9.826e-02 & 3.275e+00 & 1.723e-01 & 1.177e+02 & 3.639e-05 & 7.303e+03 & 7.754e-10 & \nodata &\nodata\\
	3.598e-01 & 1.086e-01 & 3.490e+00 & 1.508e-01 & 1.294e+02 & 2.836e-05 & 8.042e+03 & 6.033e-10 & \nodata &\nodata\\
	3.777e-01 & 1.199e-01 & 3.724e+00 & 1.319e-01 & 1.423e+02 & 2.209e-05 & 8.855e+03 & 4.693e-10 & \nodata &\nodata\\
	3.966e-01 & 1.322e-01 & 3.978e+00 & 1.151e-01 & 1.565e+02 & 1.720e-05 & 9.751e+03 & 3.651e-10 & \nodata &\nodata\\
	4.163e-01 & 1.456e-01 & 4.254e+00 & 1.002e-01 & 1.721e+02 & 1.339e-05 & 1.074e+04 & 2.839e-10 & \nodata &\nodata\\
	4.371e-01 & 1.600e-01 & 4.554e+00 & 8.683e-02 & 1.894e+02 & 1.042e-05 & 1.182e+04 & 2.208e-10 & \nodata &\nodata\\
	4.590e-01 & 1.756e-01 & 4.880e+00 & 7.502e-02 & 2.083e+02 & 8.110e-06 & 1.302e+04 & 1.717e-10 & \nodata &\nodata\\
	4.820e-01 & 1.925e-01 & 5.236e+00 & 6.465e-02 & 2.292e+02 & 6.311e-06 & 1.434e+04 & 1.335e-10 & \nodata &\nodata\\
	5.061e-01 & 2.105e-01 & 5.625e+00 & 5.552e-02 & 2.522e+02 & 4.911e-06 & 1.579e+04 & 1.037e-10 & \nodata &\nodata\\
	5.315e-01 & 2.299e-01 & 6.049e+00 & 4.752e-02 & 2.775e+02 & 3.821e-06 & 1.738e+04 & 8.062e-11 & \nodata &\nodata\\
	5.582e-01 & 2.505e-01 & 6.513e+00 & 4.056e-02 & 3.054e+02 & 2.973e-06 & 1.914e+04 & 6.265e-11 & \nodata &\nodata\\
	\enddata
	\tablecomments{Differential Intensity units: (m$^2$ s sr GV)$^{-1}$.}
\end{deluxetable*}

\begin{deluxetable*}{cccccccccc}
	\tabletypesize{\footnotesize}
	\tablecolumns{10}
	\tablewidth{0mm}
	\tablecaption{$Z=26$ -- Iron LIS\label{Tbl-IronLIS-EKin}}
	\tablehead{
		\colhead{$E_{\rm kin}$} & \colhead{Differential} &
		\colhead{$E_{\rm kin}$} & \colhead{Differential} &
		\colhead{$E_{\rm kin}$} & \colhead{Differential} &
		\colhead{$E_{\rm kin}$} & \colhead{Differential} &
		\colhead{$E_{\rm kin}$} & \colhead{Differential} \\ [-2ex] 
		\colhead{GeV/n} & \colhead{intensity} &
		\colhead{GeV/n} & \colhead{intensity} &
		\colhead{GeV/n} & \colhead{intensity} &
		\colhead{GeV/n} & \colhead{intensity} &
		\colhead{GeV/n} & \colhead{intensity}  
	}
	\startdata
	1.000e-03 & 2.434e-01 & 6.309e-02 & 2.428e+00 & 3.981e+00 & 3.246e-02 & 2.512e+02 & 1.428e-06 & 1.585e+04 & 2.969e-11\\
	1.101e-03 & 2.602e-01 & 6.948e-02 & 2.483e+00 & 4.384e+00 & 2.708e-02 & 2.766e+02 & 1.112e-06 & 1.745e+04 & 2.305e-11\\
	1.213e-03 & 2.773e-01 & 7.651e-02 & 2.531e+00 & 4.827e+00 & 2.253e-02 & 3.046e+02 & 8.663e-07 & 1.922e+04 & 1.790e-11\\
	1.335e-03 & 2.954e-01 & 8.425e-02 & 2.571e+00 & 5.315e+00 & 1.866e-02 & 3.354e+02 & 6.748e-07 & 2.116e+04 & 1.389e-11\\
	1.470e-03 & 3.148e-01 & 9.277e-02 & 2.602e+00 & 5.853e+00 & 1.540e-02 & 3.693e+02 & 5.257e-07 & 2.330e+04 & 1.079e-11\\
	1.619e-03 & 3.353e-01 & 1.022e-01 & 2.623e+00 & 6.446e+00 & 1.266e-02 & 4.067e+02 & 4.096e-07 & 2.566e+04 & 8.371e-12\\
	1.783e-03 & 3.572e-01 & 1.125e-01 & 2.635e+00 & 7.098e+00 & 1.036e-02 & 4.478e+02 & 3.192e-07 & 2.825e+04 & 6.497e-12\\
	1.963e-03 & 3.805e-01 & 1.239e-01 & 2.637e+00 & 7.816e+00 & 8.450e-03 & 4.931e+02 & 2.487e-07 & 3.111e+04 & 5.042e-12\\
	2.162e-03 & 4.052e-01 & 1.364e-01 & 2.627e+00 & 8.607e+00 & 6.864e-03 & 5.430e+02 & 1.937e-07 & 3.426e+04 & 3.912e-12\\
	2.381e-03 & 4.315e-01 & 1.502e-01 & 2.602e+00 & 9.478e+00 & 5.555e-03 & 5.980e+02 & 1.508e-07 & 3.773e+04 & 3.035e-12\\
	2.622e-03 & 4.593e-01 & 1.654e-01 & 2.564e+00 & 1.044e+01 & 4.476e-03 & 6.585e+02 & 1.175e-07 & 4.155e+04 & 2.355e-12\\
	2.887e-03 & 4.889e-01 & 1.822e-01 & 2.513e+00 & 1.149e+01 & 3.590e-03 & 7.251e+02 & 9.147e-08 & 4.575e+04 & 1.827e-12\\
	3.179e-03 & 5.202e-01 & 2.006e-01 & 2.447e+00 & 1.266e+01 & 2.870e-03 & 7.985e+02 & 7.122e-08 & 5.038e+04 & 1.417e-12\\
	3.501e-03 & 5.533e-01 & 2.209e-01 & 2.367e+00 & 1.394e+01 & 2.287e-03 & 8.793e+02 & 5.546e-08 & 5.548e+04 & 1.099e-12\\
	3.855e-03 & 5.884e-01 & 2.432e-01 & 2.270e+00 & 1.535e+01 & 1.818e-03 & 9.682e+02 & 4.318e-08 & 6.109e+04 & 8.524e-13\\
	4.245e-03 & 6.256e-01 & 2.678e-01 & 2.156e+00 & 1.690e+01 & 1.442e-03 & 1.066e+03 & 3.364e-08 & 6.727e+04 & 6.611e-13\\
	4.675e-03 & 6.648e-01 & 2.949e-01 & 2.024e+00 & 1.861e+01 & 1.141e-03 & 1.174e+03 & 2.621e-08 & 7.408e+04 & 5.127e-13\\
	5.148e-03 & 7.061e-01 & 3.248e-01 & 1.873e+00 & 2.049e+01 & 9.009e-04 & 1.293e+03 & 2.042e-08 & 8.157e+04 & 3.975e-13\\
	5.669e-03 & 7.498e-01 & 3.577e-01 & 1.705e+00 & 2.257e+01 & 7.098e-04 & 1.424e+03 & 1.591e-08 & 8.983e+04 & 3.082e-13\\
	6.242e-03 & 7.957e-01 & 3.938e-01 & 1.525e+00 & 2.485e+01 & 5.583e-04 & 1.568e+03 & 1.239e-08 & 9.892e+04 & 2.390e-13\\
	6.874e-03 & 8.440e-01 & 4.337e-01 & 1.339e+00 & 2.736e+01 & 4.385e-04 & 1.726e+03 & 9.654e-09 & \nodata &\nodata\\
	7.569e-03 & 8.947e-01 & 4.776e-01 & 1.158e+00 & 3.013e+01 & 3.440e-04 & 1.901e+03 & 7.519e-09 & \nodata &\nodata\\
	8.335e-03 & 9.479e-01 & 5.259e-01 & 9.886e-01 & 3.318e+01 & 2.695e-04 & 2.093e+03 & 5.855e-09 & \nodata &\nodata\\
	9.179e-03 & 1.004e+00 & 5.791e-01 & 8.372e-01 & 3.654e+01 & 2.110e-04 & 2.305e+03 & 4.559e-09 & \nodata &\nodata\\
	1.011e-02 & 1.062e+00 & 6.377e-01 & 7.069e-01 & 4.023e+01 & 1.650e-04 & 2.539e+03 & 3.549e-09 & \nodata &\nodata\\
	1.113e-02 & 1.122e+00 & 7.022e-01 & 5.982e-01 & 4.431e+01 & 1.290e-04 & 2.795e+03 & 2.762e-09 & \nodata &\nodata\\
	1.226e-02 & 1.186e+00 & 7.733e-01 & 5.092e-01 & 4.879e+01 & 1.008e-04 & 3.078e+03 & 2.150e-09 & \nodata &\nodata\\
	1.350e-02 & 1.251e+00 & 8.515e-01 & 4.365e-01 & 5.373e+01 & 7.859e-05 & 3.390e+03 & 1.673e-09 & \nodata &\nodata\\
	1.486e-02 & 1.319e+00 & 9.377e-01 & 3.757e-01 & 5.916e+01 & 6.124e-05 & 3.733e+03 & 1.302e-09 & \nodata &\nodata\\
	1.637e-02 & 1.389e+00 & 1.033e+00 & 3.236e-01 & 6.515e+01 & 4.770e-05 & 4.110e+03 & 1.013e-09 & \nodata &\nodata\\
	1.802e-02 & 1.462e+00 & 1.137e+00 & 2.785e-01 & 7.174e+01 & 3.714e-05 & 4.526e+03 & 7.877e-10 & \nodata &\nodata\\
	1.985e-02 & 1.536e+00 & 1.252e+00 & 2.393e-01 & 7.900e+01 & 2.891e-05 & 4.984e+03 & 6.126e-10 & \nodata &\nodata\\
	2.185e-02 & 1.612e+00 & 1.379e+00 & 2.050e-01 & 8.699e+01 & 2.250e-05 & 5.489e+03 & 4.764e-10 & \nodata &\nodata\\
	2.406e-02 & 1.689e+00 & 1.518e+00 & 1.753e-01 & 9.580e+01 & 1.751e-05 & 6.044e+03 & 3.704e-10 & \nodata &\nodata\\
	2.650e-02 & 1.767e+00 & 1.672e+00 & 1.496e-01 & 1.055e+02 & 1.363e-05 & 6.656e+03 & 2.879e-10 & \nodata &\nodata\\
	2.918e-02 & 1.846e+00 & 1.841e+00 & 1.274e-01 & 1.162e+02 & 1.060e-05 & 7.329e+03 & 2.238e-10 & \nodata &\nodata\\
	3.213e-02 & 1.925e+00 & 2.027e+00 & 1.082e-01 & 1.279e+02 & 8.250e-06 & 8.071e+03 & 1.739e-10 & \nodata &\nodata\\
	3.539e-02 & 2.003e+00 & 2.233e+00 & 9.171e-02 & 1.409e+02 & 6.419e-06 & 8.887e+03 & 1.352e-10 & \nodata &\nodata\\
	\enddata
	\tablecomments{Differential Intensity units: (m$^2$ s sr GeV/n)$^{-1}$.}
\end{deluxetable*}

\end{document}